\documentclass[aps,prd,preprint,superscriptaddress,showpacs,floatfix,
preprintnumbers,nofootinbib, longbibliography,a4paper]{revtex4-1}
\pdfoutput=1
\usepackage{color}
\usepackage{graphicx}
\usepackage{textcomp}
\usepackage{subfigure}
\usepackage{feynmp}
\usepackage{amsmath,amssymb,array}
\usepackage{enumerate}
\usepackage{hhline}
\newcommand{\mathsym}[1]{{}}

\newcommand{\s}{{\rm I}}
\newcommand{\beqa}{\begin{eqnarray}}
\newcommand{\eeqa}{\end{eqnarray}}
\newcommand{\be}{\begin{equation}}
\newcommand{\ee}{\end{equation}}
\newcommand{\ba}{\begin{array}} 
\newcommand{\ea}{\end{array}}

\begin{document} 
\vspace*{1cm}
\title{Flavour physics without flavour symmetries}
\bigskip
\author{Wilfried Buchmuller}
\email{wilfried.buchmueller@desy.de} 
\affiliation{Deutsches Elektronen-Synchrotron DESY, 22607 Hamburg, Germany}
\author{Ketan M. Patel}
\email{ketan@iisermohali.ac.in} 
\affiliation{Indian Institute of Science Education and Research Mohali, Knowledge City, Sector  81, S A S Nagar, Manauli 140306, India \vspace{1.5cm}}

%--------------------------------------------------------------
\begin{abstract}
\noindent
We quantitatively analyze a quark-lepton flavour model derived from a six-dimensional supersymmetric theory with $SO(10)\times U(1)$ gauge symmetry, compactified on an orbifold with magnetic flux. Two bulk $\mathbf{16}$-plets charged under the $U(1)$ provide the three quark-lepton generations whereas two uncharged $\mathbf{10}$-plets yield two Higgs doublets. At the orbifold fixed points mass matrices are generated with rank one or two. Moreover, the zero modes mix with heavy vectorlike split multiplets. The model possesses no flavour symmetries. Nevertheless, there exist a number of relations between Yukawa couplings, remnants of the underlying GUT symmetry and the wave function profiles of the zero modes, which lead to a prediction of the light neutrino mass scale, $m_{\nu_1} \sim 10^{-3}$ eV and heavy Majorana neutrino masses in the range from $10^{12}$ GeV to $10^{14}$ GeV. The model successfully includes thermal leptogenesis.
\end{abstract} 
%----------------------------------------------------------
\preprint{DESY 17-220}

\maketitle

\section{Introduction}
\label{sec:introduction}
The Standard Model (SM) of particle physics is a chiral gauge theory with
three copies of~a~quark-lepton generation containing a quark doublet
$q= (u,d)$, a lepton doublet $l = (\nu,e)$ and four
singlets, $u^c$, $d^c$, $e^c$ and $n^c$, of Weyl fermions in different
representations of the gauge group $G_{\rm SM} = SU(3)\times
SU(2)\times U(1)$. This gauge theory has a $U(3)^6$ flavour symmetry
which is almost completely broken by 36 complex Yukawa couplings and 6
complex Majorana mass terms. Only a $\mathbb Z_2$ matter parity and the
global $U(1)$ of baryon number survive, which is broken by an anomaly.
Most of the 84 real parameters are
unphysical and can be eliminated by a redefinition of the quark and
lepton fields, leaving 25 observables: 6 quark masses, 3 charged
lepton masses, 6 Majorana neutrino masses, 6 mixing angles in the
charged current and 4 CP violating phases. The traditional goal of
flavour physics is to reduce the number of independent input parameters 
by means of symmetries in order to obtain relations among the
various observables. These relations would then shed light on the
origin of the Yukawa couplings.

Relations between quark and lepton Yukawa matrices are obtained in
grand unified theories (GUTs) where the Standard Model gauge group is
embedded in the non-Abelian gauge groups
\(SU(4)\times SU(2)\times SU(2)\) \cite{Pati:1974yy}, \(SU(5)\) \cite{Georgi:1974sy},
\(SO(10)\) \cite{Georgi:1974my,Fritzsch:1974nn} or flipped
\(SU(5)\) \cite{Barr:1981qv,Derendinger:1983aj}. For example, in
$SU(5)$ GUTs the 36 SM Yukawa couplings are reduced to 24 couplings
and in $SO(10)$ GUTs with two Higgs ${\bf 10}$-plets only 12
independent couplings are left. However, the obtained relations
between Yukawa couplings are only partially successful and in order
to account for all measured observables one needs higher-dimensional
Higgs representations and/or higher-dimensional operators (see
  for example
  \cite{Babu:1992ia,Babu:1998wi,Matsuda:2000zp,Aulakh:2003kg,Bertolini:2004eq,Bertolini:2006pe,Grimus:2006rk,Aulakh:2008sn,Joshipura:2009tg,Altarelli:2010at,Joshipura:2011nn,Dueck:2013gca,Altarelli:2013aqa,Aulakh:2013lxa,Feruglio:2014jla,Feruglio:2015iua}
  for quantitative analyses of the fermion mass spectrum in some $SO(10)$ models). 

A partial understanding 
of the hierarchies among quark and lepton Yukawa couplings can
be obtained by means of \(U(1)\) flavour symmetries
\cite{Froggatt:1978nt} or discrete symmetries \cite{Altarelli:2010gt,King:2013eh}.
Such flavour symmetries have also been derived in string
compactifications \cite{Kobayashi:2006wq,Raby:2011jt,Nilles:2012cy,Heckman:2010bq,Abe:2014vza}. They
are of particular importance in supersymmetric compactifications where
they can forbid operators leading to proton decay.
Note, however, that none of these flavour symmetries is exact. They
are all spontaneously or explicitly broken.

Hierarchical Yukawa couplings can also be obtained in
toroidal compactifications of Super-Yang-Mills theories with magnetic
flux in ten
or fewer dimensions. The couplings between bulk Higgs and matter fields
are calculated as
overlap integrals of wave functions that have non-trivial profiles in
the magnetized extra dimensions \cite{Cremades:2004wa}. In a similar
way, Yukawa couplings of magnetized toroidal orbifolds have been analyzed
\cite{Abe:2013bca,Abe:2015yva,Matsumoto:2016okl,Fujimoto:2016zjs,Kobayashi:2016qag,Ishida:2017avx}.
The resulting flavour structure depends on the number of pairs of
Higgs doublets. In the simplest cases it appears difficult to obtain
the measured hierarchies of quark and lepton masses
\cite{Matsumoto:2016okl,Fujimoto:2016zjs}.

In this paper we pursue an alternative avenue. Our starting point is 
the six-dimensional (6D) orbifold GUT model with gauge group \(SO(10)\times
U(1)\) considered in \cite{Buchmuller:2015jna}. The GUT group $SO(10)$
is broken to different subgroups at the orbifold fixed points where
also the Yukawa couplings are generated \cite{Asaka:2001eh,Hall:2001xr}.
Abelian magnetic flux generates three  quark-lepton
families from two bulk $\mathbf{16}$-plets, $\psi$ and $\chi$, and, 
together with two uncharged $\mathbf{16}^*$-plets vectorlike split multiplets.
Moreover, the magnetic flux breaks supersymmetry \cite{Bachas:1995ik}.
Two uncharged bulk $\mathbf{10}$-plets yield two Higgs doublets. 
The 6D theory has no flavour symmetry. All quarks and leptons arise as
zero-modes of bulk $\mathbf{16}$-plets. But since their wave functions
are different, they couple with different strength to the Higgs fields
at the fixed points. As a consequence, also the effective 4D theory
has no flavour symmetries. Nevertheless, the GUT symmetry and the flux
compactification leads to a number of relations between the Yukawa 
matrices. The 36 SM complex Yukawa couplings are reduced to 12 complex
complings. In addition there are nonrenormalizable terms generating
the heavy Majorana neutrino masses and mass mixing terms between the
chiral quark-lepton generations and the vectorlike multiplets. In the
following we shall study to what extent such a structure can
quantitatively describe the measured observables, extending the 
previous work on two quark-lepton generations
\cite{Buchmuller:2017vho}.

The paper is organized as follows. In Section \ref{sec:model} we describe symmetry
breaking and zero modes of the model under consideration. Moreover,
we list the values of the zero mode wave functions at the various
fixed points and work out the Yukawa couplings which determine the
flavour spectrum. Section \ref{sec:numerical} is devoted to numerical fits of the model
to measured observables. In a first fit, light and heavy neutrino
masses and the baryon asymmetry are predicted, whereas in a second fit
the observed baryon asymmetry is also fitted. Summary and
conclusions are given in Section \ref{sec:conclusion}. Some technical features of numerical fits and results are described in Appendices \ref{app:A} and \ref{app:B}, respectively.

\section{GUT model and Yukawa couplings}
\label{sec:model}

In this section we describe  the six-dimensional
$SO(10)$ GUT model introduced in \cite{Buchmuller:2015jna}, extended 
by a pair of bulk $\mathbf{16}$-plets. This allows to account
for the flavour structure of three quark-lepton generations, with some
predictions for neutrino masses. Two additional $\mathbf{10}$-plets,
needed to cancel the 6D $SO(10)$ gauge anomalies, do not mix with
quarks and leptons and will not be discussed in the following.

The starting point is an $\mathcal N = 1$ supersymmetric 
$SO(10) \times U(1)$ gauge theory in six dimensions
with vector multiplets and hypermultiplets, compactified on the orbifold
$T^2/\mathbb Z_2$. 
One conveniently groups 6D vector multiplets into 4D vector multiplets $A =
(A_\mu, \lambda)$ and 4D chiral multiplets $\Sigma = (A_{5,6},
\lambda')$, and 6D hypermultiplets into two chiral multiplets,
$(\phi, \chi)$ and $(\phi', \chi')$ \cite{Marcus:1983wb,ArkaniHamed:2001tb}, where $(\phi',\chi')$
transform in the complex conjugate representation compared to
$(\phi, \chi)$. The origin $\zeta_\mathrm{I} =0$ is a fixed point
under reflections, $Ry=-y$, where $y$ denotes the coordinates of the
compact dimensions. Imposing chiral boundary conditions on the
orbifold, 6D $\mathcal N = 1$ supersymmetry is broken to 4D $\mathcal N = 1$
supersymmetry, and the chiral superfields $\Sigma$ and $\phi'$
are projected out.

The bulk $SO(10)$ symmetry is broken to the Standard Model group
by means of two Wilson lines. The fixed points $\zeta_i$, $i =
\mathrm{PS},\mathrm{GG},\mathrm{fl}$  are invariant
under combined lattice translations and reflection: $\hat{T}_i \zeta_i
= \zeta_i$ (see, for instance, \cite{Buchmuller:2017vho}).
Demanding that gauge fields on the orbifold satisfy the relations
\begin{equation}
\label{vectoreta}
 P_i\, A(x, \hat{T}_i y)\, P_i^{-1} =
 \eta_i\, A(x, y)\,,\quad i = \mathrm{PS}\,,\mathrm{GG}\,,
%P_\mathrm{GG} A(x, \hat{T}_\mathrm{GG} y) P_{\mathrm{GG}}^{-1} &=
% \eta_\mathrm{GG} A(x, y)\,,
%\end{split}
\end{equation}
with appropriately chosen $SO(10)$ matrices $P_i$ 
and parities $\eta_\mathrm{PS},\, \eta_\mathrm{GG} = \pm$,
the gauge group $SO(10)$ is broken to the Pati-Salam
subgroup $G_\mathrm{PS} = SU(4)\times SU(2)\times SU(2)$ and the
Georgi-Glashow subgroup $G_\mathrm{GG} = SU(5)\times U(1)_X$ at the
fixed points $\zeta_\mathrm{PS}$ and $\zeta_\mathrm{GG}$, respectively
(see Fig.~1).
In four dimensions the SM gauge group results as intersection of the
Pati-Salam and Georgi-Glashow subgroups of $SO(10)$,
$G_\mathrm{SM'} = G_\mathrm{PS} \cap G_\mathrm{GG}  
= SU(3)\times SU(2)\times U(1)_\mathrm{Y} \times U(1)_\mathrm{X}$.
Group theory implies that $SO(10)$ is broken to flipped $SU(5)$,
$G_\mathrm{fl} = SU(5)'\times U(1)_{X'}$ at $\zeta_\mathrm{fl}$.

%%%%%%%%%%%%%
\begin{figure}[t]
\centering 
\includegraphics[scale=.5]{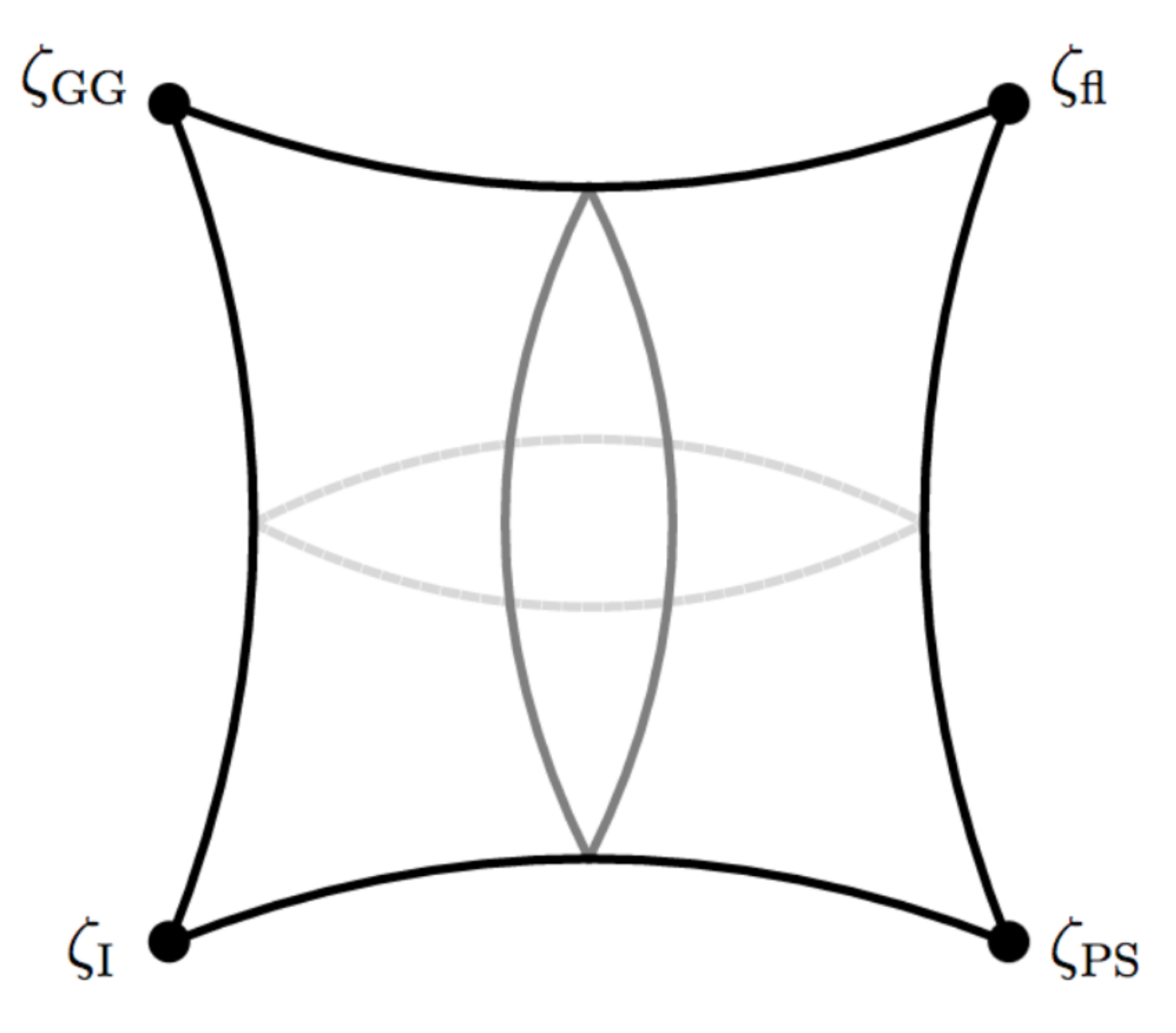}
\caption{Orbifold $T^2/{\mathbb Z_2}$  with two Wilson lines and the
fixed points $\zeta_{\rm I}$, $\zeta_{\rm PS}$, $\zeta_{\rm GG}$, and $\zeta_{\rm fl}$.}
\end{figure}
%%%%%%%%%%%%%%
Like the vector multiplets, the hypermultiplets satisfy relations
\begin{equation}
P_i\, \phi(x, \hat{T}_i y) =
 \eta_i\, \phi(x, y)\ ,\quad i = \mathrm{PS}\,,\mathrm{GG}\,,
\end{equation}
where the matrices $P_\mathrm{PS}$ and $P_\mathrm{GG}$ now depend on the
representation of the hypermultiplet (see \cite{Buchmuller:2017vho}).
The $SO(10)$ multiplets $\phi$ can be decomposed into SM
multiplets, $\phi = \{\phi^\alpha\}$.
Each of them belongs to a repesentation of
$G_\mathrm{PS}$ as well as $G_\mathrm{GG}$ and is therefore
characterized by two parities,
\begin{align}
\phi^\beta(x, \hat{T}_\mathrm{PS} y) &= \eta^\beta_\mathrm{PS}\, \phi^\beta(x, y)\,,\quad
\phi^\beta(x, \hat{T}_\mathrm{GG} y) =\eta^\beta_\mathrm{GG}\, \phi^\beta(x, y)\,.
\end{align}
They can be freely chosen subject to the requirement of
anomaly cancellations. A given set of parities then defines a 4D
model with SM gauge group. The model \cite{Buchmuller:2015jna}
contains two pairs of $\mathbf{16}$- and $\mathbf{16}^*$-plets,
$\psi$ and $\psi^c$ with parities $\eta_\mathrm{PS} = -1\,, \eta_\mathrm{GG} = +1$,
and $\Psi$ and $\Psi^c$ with parities $\eta_\mathrm{PS} = -1\,, \eta_\mathrm{GG} = -1$.
Two $\mathbf{10}$-plets contain the Higgs doublets $H_u$ and $H_d$. We
now introduce  a third pair of $\mathbf{16}$- and $\mathbf{16}^*$-plets,
$\chi$ and $\chi^c$ with parities $\eta_\mathrm{PS} = -1\,, \eta_\mathrm{GG} = -1$.

Magnetic flux is generated by a $U(1)$  background gauge field. For a bulk
$\mathbf{16}$-plet with charge $q$ and magnetic flux $f=-4\pi N/q$ 
one obtains $N$ left-handed $\mathbf{16}$-plets of zero modes. In
addition there is a split multiplet of zero modes whose quantum
numbers depend on the choice of $\eta_\mathrm{PS}$ and
$\eta_\mathrm{GG}$. We choose the charges $q=2$ and $q=1$ for $\psi$
and $\chi$, respectively, whereas $\psi^c$, $\chi^c$, $\Psi$ and
$\Psi^c$ carry zero $U(1)$ charge\footnote{We expect that charged and
  neutral $SO(10)$ singlets can be added such that all gauge and
  gravitational anomalies cancel. For the model
  \cite{Buchmuller:2015jna} this was recently shown in \cite{Buchmuller:2017wpe}. We
  also neglect the possible effect of zero modes localized at the
  fixed points, which may be needed to cancel fixed points anomalies.}. The resulting zero modes are
summarized in Table~\ref{tab1}. 
%%%%%%%%%%%%%%
\begin{table}
\begin{center}
   $\begin{array}{|c||cc|cc|cc|cc|}\hline
 \mbox{$SO(10)$} &
     \multicolumn{8}{c|}{ \mathbf{10} }
     \\ \hline
     G_\mathrm{PS} &
     \multicolumn{2}{c|}{ ( \mathbf{1}, \mathbf{2}, \mathbf{2}) } &
     \multicolumn{2}{c|}{ ( \mathbf{1}, \mathbf{2}, \mathbf{2}) } &
     \multicolumn{2}{c|}{ ( \mathbf{6}, \mathbf{1}, \mathbf{1}) } &
     \multicolumn{2}{c|}{ ( \mathbf{6}, \mathbf{1}, \mathbf{1}) }
     \\ \hline
     G_\mathrm{GG} &
     \multicolumn{2}{c|}{ \mathbf{5}^\ast{}_{-2} } &
     \multicolumn{2}{c|}{ \mathbf{5}{}_{+2} } &
     \multicolumn{2}{c|}{ \mathbf{5}^\ast{}_{-2} }  &
     \multicolumn{2}{c|}{ \mathbf{5}{}_{+2} }
     \\ \hline
 {\mathrm{parities}} &
    \eta_\mathrm{PS} & \eta_\mathrm{GG} &
    \eta_\mathrm{PS} & \eta_\mathrm{GG} &
    \eta_\mathrm{PS} & \eta_\mathrm{GG} &
    \eta_\mathrm{PS} & \eta_\mathrm{GG}
    \\ \hline 
     H_1 &
     + & - &
     + & + &
     - & - &
     - & +
     \\ 
   &  \multicolumn{2}{c|}{} & \multicolumn{2}{c|}{H_u} &
     \multicolumn{2}{c|}{} & \multicolumn{2}{c|}{}
     \\ \hline
     H_2 &
     + & + &
     + & - &
     - & + &
     - & -
     \\ 
   &  \multicolumn{2}{c|}{H_d} & \multicolumn{2}{c|}{} &
     \multicolumn{2}{c|}{} & \multicolumn{2}{c|}{}
     \\ \hline\hline
  %   H_3 &
  %   - & - &
  %   - & + &
  %   + & - &
  %   + & +
  %   \\ 
  % &  \multicolumn{2}{c|}{} & \multicolumn{2}{c|}{} &
  %   \multicolumn{2}{c|}{} & \multicolumn{2}{c|}{d}
  %   \\ \hline
  %   H_4 &
  %   - & - &
  %   - & + &
  %   + & + &
  %   + & -
  %   \\ 
 %  &  \multicolumn{2}{c|}{} & \multicolumn{2}{c|}{} &
 %    \multicolumn{2}{c|}{d^c} & \multicolumn{2}{c|}{}
 %    \\ \hline\hline
    \mbox{$SO(10)$} & \multicolumn{8}{c|}{ \mathbf{16} }
    \\ \hline
    G_\mathrm{PS} &
    \multicolumn{2}{c|}{ (\mathbf{4}, \mathbf{2}, \mathbf{1}) } &
    \multicolumn{2}{c|}{ (\mathbf{4}, \mathbf{2}, \mathbf{1}) } &
    \multicolumn{2}{c|}{ (\mathbf{4}^\ast, \mathbf{1}, \mathbf{2}) } &
    \multicolumn{2}{c|}{ (\mathbf{4}^\ast, \mathbf{1}, \mathbf{2}) }
    \\ \hline
    G_\mathrm{GG} &
    \multicolumn{2}{c|}{ \mathbf{10}_{-1} } &
    \multicolumn{2}{c|}{ \mathbf{5}^\ast{}_{+3} } &
    \multicolumn{2}{c|}{ \mathbf{10}_{-1 }} &
    \multicolumn{2}{c|}{ \mathbf{5}^\ast{}_{ +3 }, \mathbf{1}_{ -5 } }
    \\ \hline
 {\mathrm{parities}} &
    \eta_\mathrm{PS} & \eta_\mathrm{GG} &
    \eta_\mathrm{PS} & \eta_\mathrm{GG} &
    \eta_\mathrm{PS} & \eta_\mathrm{GG} &
    \eta_\mathrm{PS} & \eta_\mathrm{GG}
    \\ \hline     
\psi &
     - & + &
     - & - &
     + & + &
     + & -
     \\ 
{} &
    \multicolumn{2}{c|}{q_i}      &
    \multicolumn{2}{c|}{l_i}       &
    \multicolumn{2}{c|}{u_i^c, e_i^c}  &
    \multicolumn{2}{c|}{d_i^c, n_i^c}
    \\
{} &
    \multicolumn{2}{c|}{ }      &
    \multicolumn{2}{c|}{}       &
    \multicolumn{2}{c|}{u_3^c, e_3^c}  &
    \multicolumn{2}{c|}{}
    \\
\hline
\chi &
     - & - &
     - & + &
     + & - &
     + & +
     \\ 
{} &
    \multicolumn{2}{c|}{q_3}      &
    \multicolumn{2}{c|}{l_3}       &
    \multicolumn{2}{c|}{u_4^c, e_4^c}  &
    \multicolumn{2}{c|}{d_{3}^c, n_{3}^c}
    \\
{} &
    \multicolumn{2}{c|}{ }      &
    \multicolumn{2}{c|}{}       &
    \multicolumn{2}{c|}{}  &
    \multicolumn{2}{c|}{d_4^c, n_4^c}
    \\
\hline
     \Psi &
     - & - &
     - & + &
     + & - &
     + & +
     \\ 
{} &
    \multicolumn{2}{c|}{ }      &
    \multicolumn{2}{c|}{}       &
    \multicolumn{2}{c|}{}  &
    \multicolumn{2}{c|}{D^c,N^c}
    \\\hline\hline
  \mbox{$SO(10)$} & \multicolumn{8}{c|}{ \mathbf{16}^\ast }
    \\ \hline
    G_\mathrm{PS} &
    \multicolumn{2}{c|}{ (\mathbf{4}^\ast, \mathbf{2}, \mathbf{1}) } &
    \multicolumn{2}{c|}{ (\mathbf{4}^\ast, \mathbf{2}, \mathbf{1}) } &
    \multicolumn{2}{c|}{ (\mathbf{4}, \mathbf{1}, \mathbf{2}) } &
    \multicolumn{2}{c|}{ (\mathbf{4}, \mathbf{1}, \mathbf{2}) }
    \\ \hline
    G_\mathrm{GG} &
    \multicolumn{2}{c|}{ \mathbf{10}^\ast_{+1} } &
    \multicolumn{2}{c|}{ \mathbf{5}_{-3} } &
    \multicolumn{2}{c|}{ \mathbf{10}^\ast_{+1 }} &
    \multicolumn{2}{c|}{ \mathbf{5}_{ -3 }, \mathbf{1}_{ +5 } }
    \\ \hline
 {\mathrm{parities}} &
    \eta_\mathrm{PS} & \eta_\mathrm{GG} &
    \eta_\mathrm{PS} & \eta_\mathrm{GG} &
    \eta_\mathrm{PS} & \eta_\mathrm{GG} &
    \eta_\mathrm{PS} & \eta_\mathrm{GG}
    \\ \hline 
     \psi^c &
     - & + &
     - & - &
     + & + &
     + & -
     \\ 
{} &
    \multicolumn{2}{c|}{ }      &
    \multicolumn{2}{c|}{}       &
    \multicolumn{2}{c|}{u,e}  &
    \multicolumn{2}{c|}{}
    \\
\hline
    \chi^c &
     - & -&
     - & + &
     + & -  &
     + & +
     \\ 
{} &
    \multicolumn{2}{c|}{ }      &
    \multicolumn{2}{c|}{}       &
    \multicolumn{2}{c|}{}  &
    \multicolumn{2}{c|}{d,n}
    \\
\hline
\Psi^c &
     - & - &
     - & + &
     + & - &
     + & +
     \\ 
{} &
    \multicolumn{2}{c|}{ }      &
    \multicolumn{2}{c|}{}       &
    \multicolumn{2}{c|}{}  &
    \multicolumn{2}{c|}{D,N}
    \\
\hline
    \end{array}$
\caption{{\rm PS}- and {\rm GG}-parities for bulk $\mathbf{10}$-plets, $\mathbf{16}$-plets and $\mathbf{16}^\ast$-plets. The index $i=1,2$ labels two quark-lepton families of zero modes.}
    \label{tab1}
  \end{center}
\end{table}
%%%%%%%%%%%%%%%%%
Note that expectation values of  $N^c$ and $N$ break $U(1)_X$, and therefore $B-L$. 

The zero modes of the charged hypermultiplets have non-trivial wave
function profiles. The decomposition of all bulk $\mathbf{16}$- and $\mathbf{16}^\ast$-plets
reads
\begin{align} \label{decomposition}
%\begin{split}
\psi &= \sum_{i=1,2} \left[ q_i \psi_{-+}^{(i)}\, + \, l_i \psi_{--}^{(i)}\, + \, (d^c_i+n^c_i) \psi_{+-}^{(i)} \right] + \sum_{\alpha=1,2,3} (u^c_\alpha+e^c_\alpha) \psi_{++}^{(\alpha)}\,,  \\
\chi &= q_3 \chi_{--}^{(1)}\, + \, l_3 \chi_{-+}^{(1)}\, + \, (u^c_4+e^c_4) \chi_{+-}^{(1)}\, + \, \sum_{i=1,2}(d^c_{i+2}+n^c_{i+2}) \chi_{++}^{(i)} \,,  \\
\Psi &= D^c + N^c\,,  \quad
%\end{split}
%\begin{split}
%\Psi^c &= D+N\,, \quad
\psi^c = u+e\,,  \quad
\chi^c = d+n\,,  \quad
\Psi^c = D+N\,.
\end{align}
Here the chiral multiplet $q = (u,d)$ contains an $SU(2)$ doublet of
left-handed up- and down quarks,  $l = (\nu,e)$ a doublet of
left-handed neutrino and electron, and the charge conjugate states of
right-handed up- and down-quark, neutrino and electron are contained
in $u^c$, $d^c$, $n^c$ and $e^c$, respectively.

All Yukawa couplings and mass mixing terms depend on the values of the
wave functions at the four fixed points. For
$\psi^{(a)}_{\eta_{\mathrm{PS},\mathrm{GG}}}$ and $\chi^{(a)}_{\eta_{\mathrm{PS},\mathrm{GG}}}$ we use 
expressions given in \cite{Buchmuller:2017vho}.  For $N$ flux quanta,
a wave function $\varphi^{(a)}_{\eta_{\rm PS},\eta_{\rm GG}}(y_1,y_2)$ is given as
\begin{align} \label{wf}
\varphi^{(a)}_{\eta_{\rm PS},\eta_{\rm GG}}(y_1,y_2;N) =& {\cal N} e^{-2 \pi N y_2^2}\, \sum_{n \in \mathbb{Z}} e^{ -2 \pi N \left(n-\frac{a}{2N}\right)^2-i\pi\left(n-\frac{a}{2N}\right)(i k_{\rm PS} - k_{\rm GG})} \nonumber \\
 &\times\cos \left[ 2 \pi \left( -2nN+a+\frac{k_{\rm PS}}{2}(y_1 + i
    y_2)\right)\right]~,
\end{align}
where $\eta_{\rm PS} = e^{i \pi k_{\rm PS}}$, $\eta_{\rm GG} = e^{i
  \pi k_{\rm GG}}$ and $k_{\rm PS},k_{\rm GG}=0,1$. For $\eta_{\rm
  PS}=\eta_{\rm GG}=+1$, one gets $N+1$ massless modes with
$a=0,1,...,N$. In the remaining cases, one obtains $N$ zero modes with
$a=0,1,...,N-1$. We choose the ordering
\begin{align} \label{profile-cov}
\psi^{(2)}_{\eta_{\rm PS},\eta_{\rm GG}}(y_1,y_2) &=  \varphi^{(0)}_{\eta_{\rm PS},\eta_{\rm GG}}(y_1,y_2;2),~ \nonumber \\
\psi^{(1)}_{\eta_{\rm PS},\eta_{\rm GG}}(y_1,y_2) &=  \varphi^{(1)}_{\eta_{\rm PS},\eta_{\rm GG}}(y_1,y_2;2),~ \nonumber \\
 \psi^{(3)}_{++}(y_1,y_2) &=  \varphi^{(2)}_{++}(y_1,y_2;2),~ \nonumber \\
 \chi^{(1)}_{\eta_{\rm PS},\eta_{\rm GG}}(y_1,y_2) &=  \varphi^{(0)}_{\eta_{\rm PS},\eta_{\rm GG}}(y_1,y_2;1),~ \nonumber \\
 \chi^{(2)}_{++}(y_1,y_2) &=  \varphi^{(1)}_{++}(y_1,y_2;1).  
\end{align}
 The wave functions evaluated at the different fixed points $\zeta_\s:(y_1=0,y_2=0)$, $\zeta_{\rm PS}:(y_1=1/2,y_2=0)$, $\zeta_{\rm GG}:(y_1=0,y_2=1/2)$, $\zeta_{\rm fl}:(y_1=1/2,y_2=1/2)$ are given in Table~\ref{tab2}.

The Yukawa interactions arise at the four fixed points in the
model. Considering the unbroken symmetries and the corresponding
matter multiplets (see Table~\ref{tab1})
at the different fixed points, one obtains the following Yukawa
superpotential from the lowest-dimensional operators:\footnote{The
  magnetic flux generates a Stueckelberg mass term for the
  $U(1)$ vector boson \cite{Buchmuller:2015eya}. By means of the corresponding axion the Yukawa
  couplings can be made invariant w.r.t. the $U(1)$ symmetry.}
 \beqa \label{W_Yuk}
W_{\rm Y} & =& \delta_\s\,  \left[ \left( \frac{1}{2} y_{ua}^{\s}\, \psi \psi  + y_{ub}^{\s}\, \psi \chi + \frac{1}{2} y_{uc}^{\s}\, \chi \chi \right) H_1\, \right. \nonumber \\
& & + \left( \frac{1}{2} y_{da}^\s\, \psi \psi  +  y_{db}^\s\, \psi \chi + \frac{1}{2} y_{dc}^\s\, \chi \chi \right) H_2  \nonumber \\
&  & +  \left. \left( \frac{1}{2} y_{na}^\s\, \psi \psi  +  y_{nb}^\s\, \psi \chi + \frac{1}{2} y_{nc}^\s\, \chi \chi \right) \Psi^c \Psi^c \right] \nonumber \\
& + &  \delta_{\rm PS} \left(\frac{1}{2} y^{\rm PS}_{na}\,  4^*_\psi 4^*_\psi\, +\, y^{\rm PS}_{nb}\,  4^*_\psi 4^*_\chi\, +\,\frac{1}{2} y^{\rm PS}_{nc}\,  4^*_\chi 4^*_\chi \right)\, F F  \nonumber \\
& + &  \delta_{\rm GG} \left( \frac{1}{2} y^{\rm GG}_{ua}10_\psi 10_\psi H_5\, + y^{\rm GG}_{db} 10_\psi 5^*_\chi H_{5^*}\, + y^{\rm GG}_{\nu c} 5^*_\chi 1_\chi H_5\, +\frac{1}{2}y^{\rm GG}_{nc}1_\chi 1_\chi NN \right) \nonumber \\
& + &  \delta_{\rm fl} \left(y^{\rm fl}_{ea} \tilde{5}^*_\psi \tilde{1}_\psi H_{\tilde{5}}\, +y^{\rm fl}_{ub} \tilde{5}^*_\psi \tilde{10}_\chi H_{\tilde{5}^*}\, + \frac{1}{2}y^{\rm fl}_{dc} \tilde{10}_\chi \tilde{10}_\chi H_{\tilde{5}}\,+\frac{1}{2} y^{\rm fl}_{nc} \tilde{10}_\chi \tilde{10}_\chi \tilde{T}^*\tilde{T}^* \right)~,
 \eeqa
 where $1_\psi = n^c$ and $\tilde{1}_\chi = e^c$.
In addition to the above, the mixing between the $\psi$, $\chi$ and
$\psi^c$, $\chi^c$ at various fixed points can be written
as\footnote{Note that the structure of the mixing terms is
  considerably simpler than the one found in
  \cite{Buchmuller:2017vho}. This is due to the fact that no mixings
  with colour triplets from bulk $\mathbf{10}$-plets have to be taken
  into account to obtain satisfactory flavour mixings.}
\be \label{W_mix}
W_{\rm mix} = \sum_{p=\s,{\rm PS},{\rm GG},{\rm fl}} \delta_p \left( \mu^p_a \psi^c \psi + \mu^p_b \psi^c \chi + \mu^p_c \chi^c \chi + \mu^p_d \chi^c \psi \right)~.\ee
For simplicity, we assume universal mass terms at fixed points and set $\mu^\s_i=\mu^{\rm PS}_i=\mu^{\rm GG}_i=\mu^{\rm fl}_i \equiv \mu_i$ for $i=a,b,c,d$.
%%%%%%%%%
\begin{table}[t]
\begin{center} 
\begin{math} 
\begin{tabular}{ccccc}
\hline\hline 
  & ~~~$\zeta_{\rm I}$~~~& ~~~$\zeta_{\rm PS}$~~~& ~~~$\zeta_{\rm GG}$~~~&~~~ $\zeta_{\rm fl}$~~~\\ 
\hline
$\psi_{++}^{(\alpha)}$  & $(1.086, 1.6818,0.1454)$ & $(-1.086, 1.6818,0.1454)$ & $(1.0864, 0.1454, 1.6818)$ & $(-1.0864, 0.1454, 1.6818)$\\ 
$\psi_{+-}^{(i)}$  & $(0.7654 (1+ i), 1.6818)$ &  $(-0.7654 (1+i), 1.6818)$ & $ (0, 0)$ & $ (0, 0)$ \\ 
$\psi_{-+}^{(i)}$  & $(0.4238, 1.9546)$ & $(0,0)$ & $(1.9546, 0.4238)$ & $(0,0)$\\ 
$\psi_{--}^{(i)}$  & $(0.2749(1+i), 1.9546)$ &$(0,0)$ & $(0,0)$ & $(1.3819 (1- i), 0.3887 i)$\\ 
\hline
$\chi_{++}^{(i)}$  & $(1.4195,0.5880)$ & $(1.4195,-0.5880)$ & $(0.5880,1.4195)$ &  $(0.5880,-1.4195)$ \\ 
$\chi_{+-}^{(1)}$  & $1.4089$ &  $1.4089$ & 0 & 0 \\ 
$\chi_{-+}^{(1)}$  & $1.4089$ &  0 & $1.4089$ & 0 \\ 
$\chi_{--}^{(1)}$  & $1.2920$ & 0 & 0 & $1.2920 i$\\ 
\hline
\hline 
\end{tabular}
\end{math}
\end{center}
\caption{\label{tab2} 
Wave functions at different fixed points for one flux quantum
$N=1$. $\psi^{(\alpha)}$, $\alpha = 1,2,3$, and $\psi^{(i)}$, $i =
1,2$, are mode functions of the bulk field $\psi$ with $q=2$;
$\chi^{(i)}$, $i = 1,2$, and $\chi^{(1)}$ are mode functions of the bulk field $\chi$ with $q=1$.}
\end{table}

After the electroweak symmetry breaking the mass Lagrangian for up-type quarks obtained from Eqs.~\eqref{W_Yuk},\eqref{W_mix} can be written as
\beqa \label{up-L}
{-\cal L}_{m}^{\rm up}& = & v_u \left[ \sum_{p=\s,{\rm GG}} y^p_{ua}\, (\psi^{(i)}_{-+} \psi^{(\alpha)}_{++})|_p\, u_i u^c_\alpha + y^\s_{ub}\, (\psi^{(i)}_{-+} \chi^{(1)}_{+-})|_\s\, u_i u^c_4 \right. \nonumber \\
& + &\left.  \sum_{p=\s,{\rm fl}} y^p_{ub}\, (\chi^{(1)}_{--} \psi^{(\alpha)}_{++})|_p\, u_3 u^c_\alpha + y^\s_{uc}\, (\chi^{(1)}_{--} \chi^{(1)}_{+-})|_\s\, u_3 u_4^c \right] \nonumber \\
& + & \sum_{p=\s,{\rm PS},{\rm GG},{\rm fl}} \mu_a\, \psi^{(\alpha)}_{++}|_p\, u u^c_\alpha + \sum_{p=\s,{\rm PS}} \mu_b\, \chi^{(1)}_{+-}|_p\, u u^c_4+{\rm h.c.}~,\eeqa
where $i,j=1,2$ and $\alpha = 1,2,3$. The mass Lagrangian for the down-type quarks can be obtained from Eq.~\eqref{W_Yuk} in the same way. We obtain
\beqa \label{down-L}
{-\cal L}_{m}^{\rm down}& = & v_d \left[y^\s_{da}\, (\psi^{(i)}_{-+} \psi^{(j)}_{+-})|_\s\, d_i d^c_j + \sum_{p=\s,{\rm GG}} y^p_{db}\, (\psi^{(i)}_{-+} \chi^{(j)}_{++})|_p\, d_i d^c_{j+2} \right. \nonumber \\
& + & \left. y^\s_{db}\, (\chi^{(1)}_{--} \psi^{(i)}_{+-})|_\s\, d_3 d^c_i + \sum_{p=\s,{\rm fl}} y_{dc}^p\, (\chi^{(1)}_{--} \chi^{(j)}_{++})|_p\, d_3 d^c_{j+2} \right] \nonumber \\ 
& + & \sum_{p=\s,{\rm PS}} \mu_d\, \psi^{(i)}_{+-}|_p\, d d^c_i + \sum_{p=\s,{\rm PS},{\rm GG},{\rm fl}} \mu_c\, \chi^{(j)}_{++}|_p\, d d^c_{j+2}+{\rm h.c.}~.\eeqa
Similarly, the charged lepton mass terms are given by
\beqa \label{cl-L}
{-\cal L}_{m}^{\rm cl}& = & v_d \left[ \sum_{p=\s,{\rm fl}} y^p_{ea}\, (\psi^{(i)}_{--} \psi^{(\alpha)}_{++})|_p\, e_i e^c_\alpha + y^\s_{eb}\, (\psi^{(i)}_{--} \chi^{(1)}_{+-})|_\s\, e_i e^c_4 \right. \nonumber \\
& + &\left.  \sum_{p=\s,{\rm GG}} y^p_{eb}\, (\chi^{(1)}_{-+} \psi^{(\alpha)}_{++})|_p\, e_3 e^c_\alpha + y^\s_{ec}\, (\chi^{(1)}_{-+} \chi^{(1)}_{+-})|_\s\, e_3 e_4^c \right] \nonumber \\
& + & \sum_{p=\s,{\rm PS},{\rm GG},{\rm fl}} \mu_a\, \psi^{(\alpha)}_{++}|_p\, e e^c_\alpha + \sum_{p=\s,{\rm PS}} \mu_b\, \chi^{(1)}_{+-}|_p\, e e^c_4+{\rm h.c.}~,\eeqa
where $y^\s_{ea} = y^\s_{da}$, $y^\s_{eb} = y^\s_{db}$, $y^\s_{ec}
= y^\s_{dc}$ and $y^{\rm GG}_{eb} = y^{\rm GG}_{db}$.
For the Dirac-type neutrino
mass terms one obtains from Eq.~\eqref{W_Yuk}
\beqa \label{Dirac-L}
{-\cal L}_{m}^{\rm Dirac}& = & v_u \left[y^\s_{\nu a}\, (\psi^{(i)}_{--} \psi^{(j)}_{+-})|_\s\, \nu_i n^c_j + \sum_{p=\s,{\rm fl}} y^p_{\nu b}\, (\psi^{(i)}_{--} \chi^{(j)}_{++})|_p\, \nu_i n^c_{j+2} \right. \nonumber \\
& + & \left. y^\s_{\nu b}\, (\chi^{(1)}_{-+} \psi^{(i)}_{+-})|_\s\, \nu_3 n^c_i + \sum_{p=\s,{\rm GG}} y_{\nu c}^p\, (\chi^{(1)}_{-+} \chi^{(j)}_{++})|_p\, \nu_3 n^c_{j+2} \right] \nonumber \\ 
& + & \sum_{p=\s,{\rm PS},{\rm GG},{\rm fl}} \mu_c\, \chi^{(j)}_{++}|_p\, n n^c_{j+2} + \sum_{p=I,{\rm PS}} \mu_d\, \psi^{(i)}_{+-}|_p\, n n^c_i+{\rm h.c.}~,\eeqa
where $y_{\nu a}^\s = y_{u a}^\s$, $y_{\nu b}^\s = y_{u b}^\s$,
$y_{\nu c}^\s = y_{u c}^\s$ and $y_{\nu b}^{\rm fl} = y_{u b}^{\rm
  fl}$. Note that the mass mixing terms
$\mu_a$ and $\mu_b$ decouple one linear combination of $u^c_\alpha,
u^c_4$ and $e^c_\alpha, e^c_4$ from the low energy effective theory
whereas $\mu_c$ and $\mu_d$ decouple one linear combination of $d^c_i, d^c_{i+2}$.

The two mass mixing terms in the Dirac neutrino mass matrix for
$n,n^c_i$ and $n,n^c_{i+2}$ are comparable to the large Majorana mass
terms for  $n^c_i$ and $n^c_{i+2}$. From Eq.~\eqref{W_Yuk} one
obtains for the Majorana mass terms generated by the $B-L$ breaking
VEV  $v_{B-L} = \langle\Psi^c\rangle$: 
\beqa \label{MN-L}
{-\cal L}_{m}^{N}& = & \frac{v_{B-L}^2}{M_{P}} \left( \frac{1}{2} \sum_{p=\s,{\rm PS}} y^p_{na}\, (\psi^{(i)}_{+-} \psi^{(j)}_{+-})|_p\, n^c_i n^c_j  + \sum_{p=\s,{\rm PS}} y^p_{nb}\, (\psi^{(i)}_{+-} \chi^{(j)}_{++})|_p\, n^c_i n^c_{j+2} \right. \nonumber \\
& + &\left. \frac{1}{2} \sum_{p=\s,{\rm PS},{\rm GG},{\rm fl}}
  y^p_{nc}\, (\chi^{(i)}_{++} \chi^{(j)}_{++})|_p\, n^c_{i+2}
  n^c_{j+2} \right)+{\rm h.c.}~. \eeqa
Here $M_P = 2 \times 10^{17}$ GeV is the reduced 6D Planck scale. The eigenvalues of the corresponding $4\times 4$ matrix $M_n$ are $\mathcal{O}(v_{B-L}^2/M_P)$.
Together, Eqs.~\eqref{Dirac-L},\eqref{MN-L} yield an $8\times 8$ neutrino
mass matrix,
\be \label{nun-L}
 \mathcal{M}_{\nu,n} = \left( 
\ba{c|cc} 
0_{3 \times 3} &  v_u(Y_D)_{3 \times 4} & 0_{3\times1} \\
\hline
 v_u(Y_D^T)_{4 \times 3} & (M_n)_{4\times 4} & (\mu_D^T)_{4 \times 1} \\
0_{1 \times 3} & (\mu_D)_{1 \times 4} & 0 \\
\ea\right) \,,
\ee
where $v_u (Y_D)_{3\times 4}$ connects $\nu_i,\nu_3$ with $n_i^c,n_{i+2}^c$, and $\mu_D$
connects $n$ with $n^c_i, n^c_{i+2}$. We denote the lower right
$5\times 5$ block of the matrix by $M_N$, which has 5 Majorana mass
eigenstates. $M_D = (v_u (Y_D)_{3\times 4}, 0_{3\times 1})$ is a
$3\times 5$ Dirac neutrino mass matrix. Integrating out the five heavy
Majorana neutrinos one obtains the seesaw formula for the $3\times 3$ light neutrino mass
matrix,
\be \label{nu-L}
{M}_\nu = -{M}_D\,
{M}_N^{-1}\, {M}_D^T\,,
\ee 
from which we can extract the relevant neutrino observables.

The above mass matrices contain the complete information about the
flavour spectrum of quarks and leptons. In the following section, we
shall study in detail the viability of Eqs.~\eqref{up-L}-\eqref{nu-L} in
reproducing the experimentally observed fermion spectrum and the
predictions for neutrino masses and the baryon asymmetry via leptogenesis \cite{Fukugita:1986hr}.

It is tempting to speculate that a fit of quark and lepton mass matrices with the expressions in Eqs.~\eqref{up-L}-\eqref{nu-L} is straightforward, given the large number of free parameters. However, this is not the case since the flavour structure of the matrices is determined by the wave function profiles, with matrix elements of $\mathcal{O}(1)$, which naively is at variance with hierarchical quark and charged lepton masses. In fact, in the model \cite{Buchmuller:2017vho}, which has only one bulk 16-plet, a successful fit turned out to be impossible, despite many parameters. One quark-lepton generation always remained massless. The reason is, that before mass mixings, the mass matrices are generically rank-one. In addition, there are relations between Yukawa couplings, which reflect the different unbroken GUT groups at the different fixed points. For example, at the $SO(10)$ fixed point there are several relations, (see Eqs.~\eqref{up-L}-\eqref{Dirac-L})
\be
y^{\rm I}_{ea} = y^{\rm I}_{da}\,, \quad  y^{\rm I}_{eb} = y^{\rm I}_{db}\,, \quad y^{\rm I}_{ec} = y^{\rm I}_{dc}\,, \quad 
y^{\rm I}_{\nu a} = y^{\rm I}_{ua} \,, \quad  y^{\rm I}_{\nu b} = y^{\rm I}_{ub} \,, \quad   y^{\rm I}_{\nu c} = y^{\rm I}_{uc} \,, \ee 
and at the Georgi-Glashow and flipped $SU(5)$ fixed points one has
\be
y^\mathrm{GG}_{eb} = y^\mathrm{GG}_{db}\,, \quad 
y^\mathrm{fl}_{\nu b} = y^\mathrm{fl}_{u b}\,. \ee 
Note that the $SO(10)$ relation for $y^{\rm I}_{\nu a}$, $y^{\rm I}_{\nu b}$ and $y^{\rm I}_{\nu c}$ imply that $B-L$ has to be broken at the GUT scale in order to generate viable mass scale for the SM neutrinos.  Considering these interrelationships between the quark and lepton sectors, it is not guaranteed that one can correctly reproduce all the observables using Eqs.~\eqref{up-L}-\eqref{nu-L} despite of having substantial number of parameters.

The magnetic flux is quantized in units of the inverse volume $V_2^{-1}$ of the compact dimensions. This leads to scalar quark and lepton masses of GUT scale size \cite{Bachas:1995ik},
 \be
 m^2_{\tilde{q}} = m^2_{\tilde{l}} \sim \frac{4\pi}{V_2} 
\sim (10^{15}~{\rm GeV})^{2}\,. \ee      
An analysis of supersymmetry breaking and moduli stabilisation shows that also gravitino and gauginos are heavy (see \cite{Buchmuller:2016bgt,Buchmuller:2017vho}),
 \be
  m_{\tilde{g}} \sim m_{\tilde{W}} \sim m_{\tilde{B}} \sim m_{3/2}\,
 \sim 10^{14} ~{\rm GeV}\,. \ee
One is therefore left with an extension of the Standard Model where, depending
on radiative corrections, only two Higgs doublets and higgsinos can be light.
It is interesting that such a model can be consistent with gauge coupling
unification, which imposes constraints on $\tan\beta$ and the Higgs boson
masses \cite{Bagnaschi:2015pwa}.

The presented model assumes that all the quarks and leptons arise as zero modes
of bulk fields, caused by magnetic flux. This is the standard picture of
flux compactifications in field and string theory. Of course, in principle there
could also be ``twisted sectors", i.e. matter localized at fixed points. Matter
from bulk fields and twisted sectors has previously been considered in 
orbifold GUTs (see, for example, \cite{Asaka:2003iy}) and heterotic string compactifications
(see, for example, \cite{Buchmuller:2007qf}). However, in all these models magnetic flux has
not been included. An analysis of flux compactifications containing twisted
sectors remains a challenging question for further research.

\section{Numerical analysis of flavour spectrum}
\label{sec:numerical}
As described in the previous section, Eqs.~\eqref{up-L}-\eqref{nu-L} determine the masses and mixing parameters of the SM fermions. In order to check whether the model correctly describes the known fermion spectrum, we perform a $\chi^2$ test. For this we construct a $\chi^2$ function
\be \label{chisquare}
\chi^2 = \sum_{i=1}^n\,\left( \frac{O_i^{\rm th}(x_1,x_2,...,x_m) -
    O_i^{\rm exp}}{\sigma_i^{\rm exp}} \right)^2~, 
\ee
where $O_i^{\rm th}(x_1,x_2,...,x_m)$ are the observables estimated
from Eqs.~\eqref{up-L} - \eqref{nu-L}. They depend on the various
parameters of the model denoted as $x_j$.  The $O_i^{\rm exp}$ are the
experimentally measured values of the corresponding observables and
$\sigma_i^{\rm exp}$ are the standard deviations. As of now, 18 of
these observables are directly measured in various experiments. They
include 9 charged fermion masses, 2 neutrino mass differences, 3
mixing angles and a phase in the CKM or quark mixing matrix and 3
mixing angles in the PMNS or lepton mixing matrix
\cite{Patrignani:2016xqp}. There also exists preliminary and indirect
information about the Dirac CP phase in the lepton sector through
global fits of neutrino oscillation data 
\cite{Esteban:2016qun,Capozzi:2017ipn,deSalas:2017kay}.

The spectrum computed from Eqs.~\eqref{up-L}-\eqref{nu-L} holds at the
GUT scale. We therefore choose the GUT scale extrapolated values of the
various observables as $O_i^{\rm exp}$ for consistency. The flux
compactification also breaks supersymmetry and leads to a
two-Higgs-doublet model (2HDM) of type-II below the GUT scale
\cite{Buchmuller:2015jna}\footnote{This feature automatically suppresses the contribution of dimension-5 operators in proton decay.}.  For this reason, we use the GUT scale
values of charged fermion masses extrapolated in 2HDM with $v_u/v_d = \tan\beta
= 10$ from the latest analysis \cite{Bora:2012tx} as an example set of
data for our analysis. The effects of the
renormalization group equations (RGE) are known to be very small in
the case of the CKM parameters, and therefore we use their low scale
values from \cite{Patrignani:2016xqp}. The RGE effects
are small also in case of neutrino masses and mixing angles if the
light neutrino masses are hierarchical and follow normal
ordering. Therefore we use the low scale values of solar and
atmospheric mass squared differences and leptonic mixing angles from
the recent global fit of neutrino oscillation data performed in
\cite{Esteban:2016qun}. In order to account for RGE effects, various
threshold corrections and uncertainties due to neglecting
next-to-leading order corrections in the theoretical estimations of
flavour observables, we adopt a conservative approach and consider $30
\%$ standard deviation in the masses of light quarks (up, down and
strange) and electron and $10 \%$ standard deviation in the remaining
quantities instead of using the extrapolated experimental values of
standard deviations in Eq.~(\ref{chisquare}). Further, we assume
normal ordering for the neutrino mass spectrum. The various $O_i^{\rm
  exp}$ we use are listed in the third column of Table~\ref{resA}. 

The details of our procedure of extracting physical observables from
Eqs.~\eqref{up-L}-\eqref{nu-L} are described in Appendix \ref{app:A}.  For an
estimation of $O_i^{\rm th}$ in the case of charged fermions, we first
integrate out the heavy vectorlike states and obtain effective $3
\times 3$ matrices for each flavour. In case of neutrinos, 5 Weyl fermions,
namely $n$ in Eq.~(\ref{Dirac-L}) and $n^c_i,n^c_{i+2}$, $i=1,2$ in
Eqs.~\eqref{MN-L},\eqref{nu-L} form a $5\times 5$ Majorana mass matrix $M_N$ with
GUT scale eigenvalues. 
The mass matrix of three light neutrinos is then given by the seesaw
mass formula Eq.~(\ref{nu-L}). The various fermion masses and the CKM and
PMNS matrices are obtained using the diagonalization procedure
describe in the Appendix A. The elements of the CKM matrix are  denoted as
$V_{ij}$ while we use the PDG \cite{Patrignani:2016xqp} convention for
the parametrization of the PMNS matrix to represent its elements in terms of the mixing angles $\theta_{ij}$.

The function $\chi^2$ is numerically minimized in order to check the
viability of the model in different cases. The model contains a large
number of free parameters (20 complex couplings in 
Eq.~\eqref{W_Yuk}, 4 real mass parameters in Eq.~\eqref{W_mix} and a
real VEV $v_{B-L}$). For simplicity, we first assume that all
couplings in Eq.~\eqref{W_Yuk} are real. This leads to $m=25$ real
parameters to account for $n=19$ observed quantities.
%\footnote{Note
%  that even if $m>n$ in all the cases we discuss here, it is not
%  guaranteed that one can correctly reproduce all the observables
%  given the complicated interrelationships between the quark and
%  lepton observables as well as the complex non-linear relationship
%  between the observables and the free parameters.}. 
  We the find that
one can correctly reproduce the entire fermion spectrum with vanishing
leptonic Dirac CP phase. The reason for this can be understood as
follows. In case of real couplings in Eq.~\eqref{W_Yuk}, the CP violation
in the quark and lepton sector arise entirely from the complex
profile factors given in Table~\ref{tab2}. By choosing an appropriate
basis, it can be shown that the CP violation in the lepton sector due
to the profile factors can completely be rotated away while the same
cannot be done for the quarks. It turns out that the model can still
successfully account for the observed CP violation in the quark sector while it leads to a CP conserving
lepton sector.

The recent T2K data \cite{Abe:2017vif} and the global fits
of neutrino oscillation data show a mild preference for maximal
Dirac CP violation, $\sin \delta_{\rm MNS} \sim -1$. Moreover, in order to
account for the observed baryon asymmetry of the universe through
leptogenesis, the model would require CP phases in the lepton
sector. Motivated by this, we shall consider more general Yukawa couplings in Eq.~\eqref{W_Yuk}. Since CP violation in the quark sector is already explained without
complex couplings, we consider the minimal case in which only the Yukawa
couplings of SM singlet fermions 
are complex, i.e. $y_{na}^p,~y_{nb}^p,~y_{nc}^p$ with $p={\rm I},{\rm PS}$ and
$y_{nc}^{\rm GG},~y_{nc}^{\rm fl}$. This introduces 8 new parameters in the model. In the following, we discuss two different $\chi^2$ fits obtained for this case.

\subsection{Predicting neutrino masses and baryon asymmetry}
For the above choice of couplings the $\chi^2$ function includes $n=19$ observables as functions of $m=33$ real parameters. We minimize $\chi^2$ numerically in order to find solutions for the parameters which can reproduce the data. We find a very good fit corresponding to $\chi^2 = 0.5$ at the minimum. The results of this fit are listed in Table \ref{resA}. 
%%%%%%%%%%%%
\begin{table}[!ht]
\begin{center} 
 \begin{math} 
\begin{tabular}{cccc}
\hline
\hline 
 Observables  & $O^{\rm th}$  & $O^{\rm exp}$  & Deviations (in \%)  \\
 \hline
 $m_u$ [GeV]  & 0.00048 & 0.00048 & 0 \\
 $m_c$ [GeV]  & 0.23 & 0.23 & 0 \\
 $m_t$ [GeV]  & 74.0 & 74.1 & 0 \\
 $m_d$ [GeV]  & 0.0011 & 0.0011 & 0 \\
 $m_s$ [GeV]  & 0.018 & 0.021 & -16 \\
 $m_b$ [GeV]  & 1.19 & 1.16 & 3 \\
 $m_e$ [GeV]  & 0.00043 & 0.00044 & -2 \\
 $m_{\mu}$ [GeV]  & 0.093 & 0.093 & 0 \\
 $m_{\tau}$ [GeV]  & 1.60 & 1.61 & -1 \\
 $m^2_{\text{sol}}$ [eV$^2$]  & 0.000075 & 0.000075 & 0 \\
 $m^2_{\text{atm}}$ [eV$^2$] & 0.0025 & 0.0025 & 0 \\
 $V_{\text{us}}$ & 0.22 & 0.23 & -3 \\
 $V_{\text{cb}}$ & 0.041 & 0.041 & 0 \\
 $V_{\text{ub}}$ & 0.0036 & 0.0036 & 0 \\
 $\sin^2 \theta _{12}$ & 0.31 & 0.31 & 0 \\
 $\sin^2 \theta _{23}$ & 0.44 & 0.44 & 0 \\
 $\sin^2 \theta _{13}$ & 0.022 & 0.022 & 0 \\
 $J^Q_{\text{CP}}$ & 0.000031 & 0.000030 & 1 \\
 $\delta _{\text{MNS}}$ [$^o$] & 281 & 261 & 8 \\
 \hline
 \hline
 \multicolumn{4}{c}{Predictions} \\
 \hline
 $\alpha _{21}$ [$^o$] & 273 & $M_{N_1}$ [GeV]  & $1.8 \times 10^{10}$ \\
 $\alpha _{31}$ [$^o$] & 215 & $M_{N_2}$ [GeV]  & $6.3\times 10^{10}$ \\
 $m_{\nu _1}$ [eV] & 0.0043 & $M_{N_3}$ [GeV]  & $1.1\times 10^{11}$ \\
 $m_{\beta \beta}$  [eV]  & 0.0004 & $M_{N_4}$ [GeV]  & $1.7\times 10^{12}$ \\
$ m_{\beta }$  [eV]  & 0.0098 & $M_{N_5}$ [GeV]  & $2.7\times 10^{13}$ \\
 $\eta _B$ & $ 5.2 \times 10^{-12}$ & &  \\
\hline
\hline 
\end{tabular}
 \end{math}
\end{center}
\caption{\label{resA} Fit without leptogenesis: the results obtained for the best fit corresponding to $\chi^2 = 0.5~$.} 
\end{table}
%%%%%%%%%%%%%
It is remarkable that all observables are fitted to their experimental
values with very small deviations. The maximum deviation is found in
the  strange quark mass which is still smaller than the allowed
$30\%$ deviation from its experimental value extrapolated at the GUT scale. The fitted values of
parameters are listed in Appendix \ref{app:B}. 

At the bottom of Table \ref{resA} we show predictions for various
quantities that can be estimated from the fitted values of the
parameters. These include the Majorana phases ($\alpha_{21}, \alpha_{31}$), the
mass of the lightest SM neutrino $m_{\nu_1}$, the effective
neutrinoless double beta decay mass $m_{\beta \beta}$, the mass
measured in standard beta decay $m_{\beta}$ and the masses of the heavy neutrinos $M_{N_\alpha}$ with $\alpha=1,...,5$. 
As a comparison with the subsequent fit will show, the order of
magnitude of the absolute neutrino mass scale, i.e. $m_{\nu_1}$, is a robust prediction whereas the remaining quantities
can change significantly if the fit is slightly varied.

The baryon asymmetry generated by decays and inverse decays of the
lightest singlet neutrino can be written as \cite{Buchmuller:2004nz,Biondini:2017rpb}
\be \label{etaB1}
\eta_B = 0.96 \times 10^{-2}\,  \epsilon_1\,  \kappa_f\, ,
\ee
where the CP asymmetry is given by \cite{Covi:1996wh}
\be \label{eps1}
\epsilon_1 = -\frac{3}{16 \pi\, \tilde{m}_1} {\rm Im}\left[
  (h^\dagger M_\nu h^*)_{11}\right]\,,
\ee
and washout processes are taken into account by the efficiency
factor
\be\label{kappa1}
\kappa_f \simeq 2 \times 10^{-2} \times \left( \frac{0.01 ~{\rm eV}}{\tilde{m}_1}\right)^{1.1}\,.
\ee
CP asymmetry and washout processes depend on the effective neutrino mass
\be \label{mtilde1}
\tilde{m}_1 = \frac{v_u^2}{M_{N_1}} \left(h^\dagger h\right)_{11}\,.
\ee
In Eqs.~\eqref{eps1} and \eqref{mtilde1}, $h$ denotes the Dirac neutrino Yukawa
matrix in a basis where the mass matrix of the heavy neutrinos is
diagonal, i.e. $h= Y_D U_N$ with $U_N^T M_N U_N = {\rm
  diag.}(M_{N_1},...,M_{N_5})$. In order to obtain the expression
\eqref{eps1} for the CP asymmetry, a summation over lepton flavours in
the final state has to be carried out.

Using the parameters of the fit, one obtains for the baryon asymmetry
generated from $N_1$,
$\eta_B \simeq 5.2 \times 10^{-12}$, which is two orders of magnitude smaller than
the observed value $\eta_B \simeq (6.10 \pm 0.04) \times 10^{-10}$ \cite{Patrignani:2016xqp}.
However, for the heavy Majorana masses given in Table~\ref{resA}, the
baryon asymmetry calculated from Eqs.~\eqref{etaB1}-\eqref{kappa1}
can be modified by flavour effects of charged leptons and other heavy
neutrinos by more than an order of magnitude
\cite{Blanchet:2006ch,Dev:2017trv}. To obtain a realistic estimate of
the baryon asymmetry, the flavour effects of charged leptons and in
particular the contributions of the heavier Majorana neutrinos have to
be taken into account.

From Eqs.~\eqref{nu-L} and \eqref{etaB1}-\eqref{mtilde1} one can
easily read off how a rescaling of couplings may lead to a baryon
asymmetry enlarged by two orders of magnitude. Rescaling $h$ by a
factor 10 while keeping the neutrino masses constant, i.e. rescaling
$M_N$ by a factor 100, enhances $\epsilon_1$ by a factor 100, leaving
$\tilde{m}_1$ and $\kappa_f$ unchanged. Hence, $\eta_B$ is indeed
enlarged by a factor 100. It is not clear, however, whether such a
rescaling can be made consistent with a description of the quark
sector since the Dirac neutrino Yukawa couplings and the up-quark
Yukawa couplings are related.

\subsection{Predicting neutrino masses}
We now perform a fit including the baryon asymmetry $\eta_B$ in the $\chi^2$ function in
order to check the viability of model in reproducing the correct
baryon asymmetry together with the flavour spectrum. The number of input parameters are same as the before. The results are displayed in Table \ref{resB}. 
%%%%%%%%%%%%
\begin{table}[!ht]
\begin{center} 
 \begin{math} 
\begin{tabular}{cccc}
\hline
\hline 
 Observables  & $O^{\rm th}$  & $O^{\rm exp}$  & Deviations (in \%)  \\
 \hline
 $m_u$ [GeV] & 0.00048 & 0.00048 & 0 \\
 $m_c$ [GeV] & 0.23 & 0.23 & 0 \\
 $m_t$ [GeV] & 74.1 & 74.1 & 0 \\
 $m_d$ [GeV] & 0.00096 & 0.00113 & -15 \\
 $m_s$ [GeV] & 0.018 & 0.021 & -18 \\
 $m_b$ [GeV] & 1.16 & 1.16 & 0\\
 $m_e$ [GeV] & 0.00051 & 0.00044 & 16 \\
 $m_{\mu}$ [GeV] & 0.094 & 0.093 & 1 \\
 $m_{\tau}$ [GeV] & 1.61 & 1.61 & 0 \\
 $m^2_{\text{sol}}$  [eV$^2$] & 0.000075 & 0.000075 & 0 \\
 $m^2_{\text{atm}}$  [eV$^2$] & 0.0025 & 0.0025 & 0 \\
 $V_{\text{us}}$ & 0.23 & 0.23 & 0 \\
 $V_{\text{cb}}$ & 0.041 & 0.041 & 0 \\
 $V_{\text{ub}}$ & 0.0035 & 0.0035 & 0 \\
 $\sin^2 \theta_{12}$ & 0.31 & 0.31 & 0 \\
 $\sin^2 \theta_{23}$ & 0.44 & 0.44 & 0 \\
 $\sin^2 \theta_{13}$ & 0.022 & 0.022 & 0 \\
 $J^Q_{\text{CP}}$ & 0.000030 & 0.000030 & 0 \\
 $\delta _{\text{MNS}}$ [$^o$] & 279 & 261 & 7 \\
 $\eta _B$ & $6.1 \times 10^{-10}$ & $6.1 \times 10^{-10}$ & 0 \\
 \hline
 \hline
 \multicolumn{4}{c}{Predictions} \\
 \hline
 $\alpha _{21}$ [$^o$] & 129 & $M_{N_1}$ [GeV] & $1.3\times 10^{12}$ \\
 $\alpha _{31}$ [$^o$] & 353 & $M_{N_2}$ [GeV] & $2.0\times 10^{14}$ \\
 $m_{\nu _1}$ [eV] & 0.0017 & $M_{N_3}$ [GeV] & $3.5\times 10^{14}$ \\
 $m_{\beta \beta}$ [eV] & 0.0026 & $M_{N_4}$ [GeV] & $3.7\times 10^{14}$ \\
 $m_{\beta}$ [eV] & 0.0089 & $M_{N_5}$ [GeV] & $4.6\times 10^{14} $ \\
\hline
\hline 
\end{tabular}
 \end{math}
\end{center}
\caption{\label{resB} Fit with leptogenesis: the results obtained for the best fit corresponding to $\chi^2 = 0.95~$.} 
\end{table}
%%%%%%%%%%%%%
We obtain the minimal $\chi^2 = 0.95$ which is slightly higher
compared to the previous case but it can be still considered a very
good fit. The resulting input parameters are listed in Appendix \ref{app:B}.

Compared to the first fit the Majorana phases $\alpha_{21}$ and
$\alpha_{31}$ have changed by about 50\%. The order of magnitude of
the light neutrino masses has remained the same whereas the heavy
neutrino masses have increase by two orders of magnitude, as
expected. Correspondingly, the $B-L$ breaking VEV increases
by a factor 10.
The increase of the heavy Majorana masses has the interesting effect that the baryon asymmetry is
now indeed dominated by decays and inverse decays of the Majorana
neutrino $N_1$. Since $M_{N_2} \ldots M_{N_5} \sim
10^{14}~\mathrm{GeV}$, they are likely not to be produced from the
thermal bath and therefore they have no effect on the baryon
asymmetry. Moreover, the enhanced mass $M_{N_1} \sim
10^{12}~\mathrm{GeV}$ now lies in the unflavoured regime where flavour effects
of charged leptons can be neglected. For the effective light neutrino
mass we find
\be
\tilde{m}_1 = 0.023\, \mathrm{eV}\,,
\ee
lying precisely in the mass range
\be
\sqrt{m^2_{\text{sol}}} < \tilde{m}_1 < \sqrt{m^2_{\text{atm}}} \,.
\ee
Hence, leptogenesis takes place in the preferred strong washout regime
where the final asymmetry is independent of initial conditions. For
this value of $\tilde{m}_1$ the heavy Majorana neutrino mass has to
satisfy the lower bound $M_1 > 10^{11}\mathrm{GeV}$ (see Fig.~10 in
\cite{Buchmuller:2004nz}), which is also satisfied.
We conclude that the estimation of the baryon asymmetry and the
fit to the fermion spectrum are self-consistent. 

It is instructive to reconstruct from the fitted values of the input
parameters given in Table \ref{parameters} how the description of the flavour
spectrum and baryogenesis is accomplished. The mixing of the zero
modes of $\psi$ and $\chi$  via the heavy vectorlike multiplets is 
difficult to disentangle but it is clear that largest up-type and
down-type Yukawa couplings scale as one expects for the heaviest
generation,
\be
y^{\s}_{uc}\ \sim\ y^{\rm fl}_{ub}\ \sim\ \frac{m_t}{m_b}
\frac{y^{\s}_{dc}}{\tan\beta}\ \sim  \frac{m_t}{m_b}
\frac{y^{\rm fl}_{dc}}{\tan\beta} \,.
\ee 
Very important are also the relations at the $SO(10)$ fixed point
$y^\s_{ea} = y^\s_{da}$ and $y^\s_{\nu c} = y^\s_{uc}$ (see Eqs.~\eqref{cl-L},
\eqref{Dirac-L}). The last one implies that $B-L$ is broken at the GUT
scale and therefore $m_{\nu_1} \sim 10^{-3}\,  \mathrm{eV}$.

The Yukawa couplings vary over a range comparable to the range in the
Standard Model. This, together with mass mixings with vectorlike states
and wave function values differing by an order of magnitude leads to a
successful fit of the measured observables.

\section{Summary and conclusions}
\label{sec:conclusion}

Six-dimensional supersymmetric theories with GUT gauge symmetries are
an attractive intermediate step towards embedding the Standard Model
in string theory. We have analyzed the structure of Yukawa couplings
and mass mixings that occur in an orbifold compactification of a 6D
$SO(10)$ GUT model with Abelian magnetic flux. Three quark-lepton
generations are generated as zero modes of bulk ${\bf 16}$-plets together with
two Higgs doublets obtained from two bulk ${\bf 10}$-plets and further
vectorlike split multiplets. Although all quarks and leptons have the
same origin, they have different wave functions in the compact
dimensions and therefore different couplings to the Higgs fields at
the orbifold fixed points.

The underlying GUT symmetry and the wave function
profiles of the zero modes imply a number of relations between the
various Yukawa couplings. In a minimal setup the model has 33 real
parameters. It is non-trivial that a good fit is possible to quark and lepton
masses and mixings, CP violating phases and the baryon asymmetry
via leptogenesis (20 observables). Due to  $SO(10)$ relations between
up-quark and Dirac neutrino Yukawa couplings, $B-L$ is broken at the
GUT scale. The smallest neutrino mass is predicted to be   
$m_{\nu_1} \sim 10^{-3}\, \mathrm{eV}$ and also the neutrino masses
$m_\beta$ and $m_{\beta\beta}$, to be measured in standard
beta decay and neutrinoless double beta decay, are very small.
Heavy Majorana neutrino masses are predicted in the range from
$10^{12}\, \mathrm{GeV}$ to $10^{14}\, \mathrm{GeV}$, and the effective
light neutrino mass is $\tilde{m}_1 = 0.023\ \mathrm{eV}$. Hence, the baryon
asymmetry is indeed dominated by decays and inverse decays of the
lightest GUT scale Majorana neutrino and flavour effects on the
generated asymmetry are negligible. It is remarkable that all light
neutrino masses lie in the neutrino mass window $10^{-3}\, \mathrm{eV} <
m_{\nu_i} < 0.1\, \mathrm{eV}$ where thermal leptogenesis works best.

The model presented in this paper addresses the question of flavour physics in
flux compactifications, but it is incomplete in several respects. First of all, the vacuum expectation values $\langle H_u\rangle$, $\langle H_d\rangle$ and $\langle N\rangle$ correspond to flat directions of the model. Hence, the  determination of the scales of electroweak breaking and $B-L$ breaking require further interactions and parameters which remain to be specified. Another important point concerns the effect of the large mass mixing terms on the zero mode profiles (for a recent discussion, see \cite{Ishida:2017avx}). In principle, one has to analyse numerically the differential equations for the bulk wave functions including the mixing terms. This may lead to $\mathcal{O}(1)$ effects on the wave functions at the fixed points. However, since the values of the wave functions at fixed points are already $\mathcal{O}(1)$, we expect no qualitative change of our discussion, but rather a quantitative change in the numerical values of the free parameters. These questions will be studied in detail in a future analysis.

Our results provide a non-standard perspective on the flavour
problem. Traditionally, one searches for flavour symmetries to
understand the hierarchies of fermion masses and mixings. In the
considered model with flux compactification the quarks and leptons of
the three generations have different internal wave functions and
therefore different couplings to the Higgs fields. As a consequence, there is no
fundamental flavour symmetry. The effective 6D theory still contains
unexplained Yukawa couplings which may be related to geometry and
fluxes if the orbifold singularities are resolved in a ten-dimensional
theory. The presented model illustrates that in string compactifications flavour symmetries are not
fundamental, although they may occur as approximate accidental
symmetries in specific compactifications.

\section*{Acknowledgements}
We thank Pasquale di Bari, Markus Dierigl, Paul Oehlmann, Yoshiyuki Tatsuta, Daniel Wyler and Tsutomu Yanagida for valuable discussions. This work was supported by the German Science Foundation (DFG) within the Collaborative Research Center (SFB) 676 ``Particles, Strings and the Early Universe''. The work of KMP  was partially supported by SERB Early Career Research Award (ECR/2017/000353) and by a research grant under INSPIRE Faculty Award (DST/INSPIRE/04/2015/000508) from the Department of Science and Technology, Government of India. Computational work was carried out using HPC cluster facility of IISER Mohali. KMP thanks the DESY Theory Group for the kind hospitality during the initial stage of this work.

\section*{Appendix}
\begin{appendix}
\section{Extraction of masses and mixing parameters}
\label{app:A}
In this appendix 
we discuss our method of extracting physical observables from
Eqs.~\eqref{up-L}-\eqref{nu-L}. For the charged fermions, $f=u,d,e$,
Eqs.~\eqref{up-L}, \eqref{down-L} and \eqref{cl-L} can generally be written as
\be  \label{Lf}
{-\cal L}_{m}^f = \left( \ba{cccc} f_1 & f_2 & f_3 & f \ea \right) M_f \left( \ba{c} f^c_1 \\ f^c_2\\ f^c_3\\ f^c_4 \ea \right)+{\rm h.c}~, \ee where
\be 
M_f= \left( \ba{cccc} \multicolumn{4}{c}{v_f\, Y_f} \\ \mu^f_\alpha \ea \right) ~,  \ee
$v_e=v_d = v \cos\beta$, $v_u=v \sin\beta$ and $v=174$ GeV. $Y_f$ is a $3 \times 4$ Yukawa coupling matrix and $\mu^f_\alpha$, $\alpha = 1,...,4$, are the GUT scale mass mixing terms. We then obtain a hermitian matrix
\be \label{Hf}
H_f \equiv M_f\, M_f^\dagger= \left( \ba{c|c}v_f^2\, Y_f\,Y_f^\dagger & v_f\, (Y_f)_{i\alpha}\mu^*_\alpha \\
\hline v_f\,  (Y^*_f)_{i\alpha}\mu_\alpha & \tilde{\mu}^2_f \ea \right)~,  \ee
with $\tilde{\mu}^2_f = \sum_{\alpha} |\mu^f_\alpha|^2$. One
typically finds $(H_f)_{44} \gg (H_f)_{i4} \gg (H_f)_{ij}$ with
$i=1,2,3$. One linear combination of $f_1$, $f_2$ and $f_3$ forms
together with $f$ a
Dirac fermion with GUT scale mass and decouples from the low
energy spectrum. After integrating it out, we obtain an
effective $3\times 3$ matrix ${\tilde{H}}_f$ for the three 
families of SM fermions,
\beqa
({\tilde{H}}_f)_{ij} &=& v_f^2 (Y_f Y_f^\dagger)_{ij} -
\frac{1}{\tilde{\mu}^2_f} (H_f)_{i4} (H_f^*)_{j4} \nonumber \\
 &=& v_f^2 (Y_f Y_f^\dagger)_{ij} - v_f^2 (Y_f)_{i\alpha}
 (Y_f^*)_{j\beta} \frac{\mu_\alpha^{f*} \mu_\beta^f}{\tilde{\mu}^2_f}\,.
\eeqa

In case of the three families of light neutrinos we similarly construct $\tilde{H}_\nu = M_\nu M_\nu^\dagger$ using the $3 \times 3$ Majorana neutrino mass matrix $M_\nu$ obtained from Eq.~\eqref{nu-L}. The hermitian matrices $\tilde{H}_f$ obtained for $f=u,d,e,\nu$ are then diagonalized using $U_f^\dagger \tilde{H}_f U_f = {\rm Diag.}(m_{f_1}^2,m_{f_2}^2,m_{f_3}^2)$ where $m_{f_i}$ are the physical masses of corresponding fermions. The CKM and PMNS mixing matrices are constructed using $V = U_u^\dagger U_d$ and $U = U_l^\dagger U_\nu$, respectively.

The effective masses for standard beta decay and neutrinoless double beta decay denoted by $m_{\beta}$ and $m_{\beta \beta}$, respectively, are obtained using
\be \label{onbb}
m_\beta = \sqrt{(M_{\nu f}\, M_{\nu f}^\dagger)_{ee}}~~{\rm
  and}~~m_{\beta \beta} = |(M_{\nu f})_{ee}|\,,
\ee
where $M_{\nu f}$ is the neutrino mass matrix in the diagonal basis of charged leptons and is given by $M_{\nu f} = U_l^\dagger M_\nu U_l^*$.

\section{Fitted values of parameters}
\label{app:B}
We list here the values of input parameters of the model defined in Eqs. (\ref{W_Yuk},\ref{W_mix}) obtained from the two fits. The GUT scale mixing parameters $\mu_{a,b,c,d}$ are given in the unit of the reduced Plank scale, $M_P = 2 \times 10^{17}$ GeV.
%%%%%%%%%%%%
\begin{table}[t]
\begin{center} 
 \begin{math} 
\begin{tabular}{ccc}
\hline
\hline 
 Parameters  & Fit 1 (Table \ref{resA})  & Fit 2 (Table \ref{resB})  \\
 \hline
$y_{ua}^\s$ & $-0.5115 \times 10^{-3}$ &  $0.1148 \times 10^{-2}$ \\
$y_{da}^\s$ & $0.4472 \times 10^{-4}$ &  $-0.1314 \times 10^{-3}$ \\
$y_{na}^\s$ & $(0.8090 +0.9252\, i) \times 10^{-3}$ & $(0.2728 + 0.0583\, i)\times 10^{-3}$ \\
$y_{ub}^\s$ & $0.7538 \times 10^{-2}$ &  $0.2691 \times 10^{-1}$\\
$y_{db}^\s$ & $-0.2982 \times 10^{-2}$ &  $0.8655 \times 10^{-3}$\\
$y_{nb}^\s$ & $(-0.0829 -0.3117\, i) \times 10^{-2}$ & $ (-0.6072 + 0.6514\, i)\times 10^{-2}$\\
$y_{uc}^\s$ & $-0.2341$ &  $-0.2311$\\
$y_{dc}^\s$ & $0.4630 \times 10^{-1}$ &  $-0.3160 \times 10^{-1}$\\
$y_{nc}^\s$ & $-0.4134-0.6924\, i$ &  $(-0.3838 + 0.2643\, i)\times 10^{-1}$\\
$y_{ua}^{\rm GG}$ & $-0.9908 \times 10^{-6}$ & $-0.9914 \times 10^{-6}$ \\
$y_{db}^{\rm GG}$ & $0.1278 \times 10^{-3}$ &  $0.1186 \times 10^{-3}$ \\
$y_{\nu c}^{\rm GG}$ & $0.2355 \times 10^{-1}$ &  $0.2220$ \\
$y_{nc}^{\rm GG}$ & $(0.0951 -0.1447\, i)\times 10^{-1}$ & $0.1422 - 0.0589\, i$ \\
$y_{ea}^{\rm fl}$ & $0.1756 \times 10^{-2}$ & $-0.2330 \times 10^{-1}$ \\
$y_{ub}^{\rm fl}$ & $-0.1058 \times 10^{-1}$ & $ -0.1616$\\
$y_{dc}^{\rm fl}$ & $-0.5899 \times 10^{-2}$ &  $0.1149 \times 10^{-1}$\\
$y_{nc}^{\rm fl}$ & $(0.2005 -0.2579 \, i) \times 10^{-1}$ &  $(-0.3909 + 0.9498\,i)\times 10^{-2}$\\
$y_{na}^{\rm PS}$ & $(-0.0362+0.2929 \, i)\times 10^{-3}$ & $(0.0469 + 0.4710\, i)\times 10^{-3}$ \\
$y_{nb}^{\rm PS}$ & $(0.4811 +0.2618 \, i)\times 10^{-2}$ &  $(0.3124 + 0.5919\, i)\times 10^{-2}$\\
$y_{nc}^{\rm PS}$ & $(0.2737 +0.0246\, i)\times 10^{-1}$ &  $(0.8301 - 0.3691\, i)\times 10^{-1}$\\
$\mu_a$ & $0.9625$ $M_P$ & $0.1095\times 10^{-3}$  $M_P$\\
$\mu_b$ & $0.2191 \times 10^{-2}$ $M_P$ & $0.3401\times 10^{-2}$ $M_P$ \\
$\mu_c$ & $0.2228 \times 10^{-4}$ $M_P$ & $ 0.7488 \times 10^{-2}$ $M_P$\\
$\mu_d$ & $0.2071\times 10^{-4}$ $M_P$ &  $0.9124\times 10^{-1}$ $M_P$\\
$v_{B-L}$ & $0.8360\times 10^{-2}$ $M_P$ &  $0.8522\times 10^{-1}$ $M_P$\\
\hline
\hline 
\end{tabular}
 \end{math}
\end{center}
\caption{\label{parameters} The fitted values of input parameters obtained for the Fit 1 and Fit 2 displayed in Table \ref{resA} and Table \ref{resB}, respectively.} 
\end{table}
%%%%%%%%%%%%%

\end{appendix}
%%%

%\bibliography{references.bib}
\bibliography{references}
\end{document}